**Assessing cognitive function among older adults using machine learning and wearable device data: a feasibility study**

Collin Sakal, MSc[1]; Tingyou Li[1], Juan Li, PhD[2]; Xinyue Li, PhD[1*]

**Affiliations**

1. School of Data Science, City University of Hong Kong, Hong Kong SAR, China
2. Center on Aging Psychology, Key Laboratory of Mental Health, Institute of Psychology, Chinese Academy of Sciences, Beijing, China

***Address correspondence to:** Xinyue Li, PhD
Xinyue Li, PhD
School of Data Science, City University of Hong Kong, Hong Kong SAR, China
Postal address: 83 Tat Chee Avenue, Lau-16-224, Kowloon Tong, Hong Kong SAR, China
Telephone number: (+852)34422180
Email: xinyueli@cityu.edu.hk

## Abstract

Timely implementation of interventions to slow cognitive decline among older adults requires accurate monitoring to detect changes in cognitive function. Data gathered using wearable devices that can continuously monitor factors known to be associated with cognition could be used to train machine leaning models and develop wearable-based cognitive monitoring systems. Using data from over 2,400 older adults in the National Health and Nutrition Examination Survey (NHANES) we developed prediction models to differentiate older adults with normal cognition from those with poor cognition based on outcomes from three cognitive tests measuring different domains of cognitive function. During repeated cross validation CatBoost, XGBoost, and Random Forest models performed best when predicting cognition based on processing speed, working memory, and attention (median AUCs ≥0.82) compared to immediate and delayed recall (median AUCs ≥0.72) and categorical verbal fluency (median AUC ≥0.68). Activity and sleep parameters were also more strongly associated with processing speed, working memory, and attention compared to other cognitive subdomains. Our work provides proof of concept that wearable-based cognitive monitoring systems may be a viable alternative to traditional methods for monitoring processing speeds, working memory, and attention. We further identified novel metrics that could be targets in future causal studies seeking to better understand how sleep and activity parameters influence cognitive function among older adults.

# Introduction

Cognitive function naturally wanes as a result of aging, but cognitive decline beyond what is biologically normal can lead to a loss of independence and lower quality of life[1,2]. Interventions exist that can preserve cognition among older adults[3,4], but the effectiveness of such interventions hinges on accurate monitoring to detect changes in cognitive function for timely implementation. Existing methods used for monitoring cognition are limited in their ability to repeatedly measure cognitive capabilities across time. For example, standardized cognitive tests are time consuming and require a trained professional to administer. Such tests additionally have a limited number of validated versions, meaning they cannot be used repeatedly over short timeframes due to test familiarity resulting in biased measurements that don't reflect underlying changes in cognitive function. Machine learning-based tools have also been developed, but many require expensive and labor-intensive data collection procedures such as gathering genetic biomarkers, conducted magnetic resonance imaging (MRI), or administering lengthy surveys[5-9]. Novel approaches are therefore needed that can repeatedly assess cognitive function among older adults to overcome the test-retest reliability, time, and labor limitations of existing methods.

Wearable devices offer a convenient mechanism for collecting data that could be used to monitor cognition using machine learning models, however, it is currently unclear if such data is sufficient to accurately assess cognitive function. Previous studies have shown that older adults with poor cognition have worse sleep quality, sleep insufficiently many hours per night, are more sedentary, and have different circadian rhythm characteristics relative to older adults with normal cognition[10-14]. All such metrics, namely sleep, activity, and circadian parameters can be captured using accelerometers embedded in wearable devices. Other wearable-device metrics that can manually be collected through user interfaces such demographic and lifestyle factors are also associated with cognition[10]. While wearable-device data has been used extensively to identify factors associated with cognitive function, the application of such data for prediction tasks has been limited to small samples of participants with existing impairments. For example, Bringas et al. used deep learning to classify Alzheimer's disease stages using accelerometry[15], Mc Ardle et al. used accelerometer data from gait assessments to differentiate between dementia subtypes[16], and Rykov et al. used wearable device data and machine learning assess cognition among older adults with mild cognitive impairments[17]. While results from existing studies are promising, cognitive monitoring systems based on wearable device data cannot be used in the general population until it is determined if prediction models trained on such data can accurately differentiate between older adults with normal cognition and those who may require further evaluation for potential interventions. Examining predictive ability across different measures of cognitive function could also uncover which subdomains of cognition are best suited to be monitored using wearable-based prediction models.

Using data from over 2,400 older adults in the National Health and Nutrition Survey[18] (NHANES) we quantified the ability of multiple machine learning models trained on wearable device data to predict cognitive function based on three tests that assess immediate, delayed, and working memory, categorical verbal fluency, processing speed, and sustained attention: the Digit Symbol Substitution Test (DSST), the Consortium to Establish a Registry for Alzheimer's Disease Word-Learning subtest (CERAD-WL), and the Animal Fluency Test (AFT). Associations between sleep, activity, circadian rhythm, and ambient light exposure metrics with each cognitive outcome were further examined to uncover possible targets for future studies seeking to identify causal factors relating to cognitive impairments.

## Results

### Cohort Characteristics

A total of 2,479 older adults with an average age of 69.6 (51.7% male) were included in this study (Table 1). Three categorizations of poor cognitive function were constructed across different subdomains. A total of 625 (25.2%) individuals were categorized as having poor cognition based on the processing speed, working memory, and attention based on Digit Symbol Substitution Test (DSST) scores, 609 (24.6%) based on immediate and delayed memory using Consortium to Establish a Registry for Alzheimer's Disease Word-Learning subtest (CERAD-WL) results, and 587 (23.7%) based on categorical verbal fluency derived from the Animal Fluency Test (AFT). Participants with poor cognition were significantly less active during daytime hours across all categorizations (Figure 1). Cohort characteristics stratified by poor cognition based on each test can be found in Supplementary Tables 1-3.

### Wearable device data best predicts poor cognition based on processing speed, working memory, and attention compared to other subdomains

Ten features were included in each model as no significant improvements in predictive performance were observed for larger feature sets during the forward feature selection procedure (Figure 2). During repeated cross validation the CatBoost models had the highest median AUC compared to the XGBoost and Random Forest models when predicting poor cognition based on processing speed, working memory, and attention (median AUC: 0.84, IQR: 0.81 to 0.85), immediate and delayed memory (median AUC: 0.73, IQR: 0.71 to 0.76), and categorical fluency (median AUC: 0.71, IQR 0.68 to 0.72) as shown in Figure 3 and Table 2. CatBoost also showed the highest median AUPRC when predicting poor cognition based on processing speed, working memory, and attention (median AUPRC: 0.62, IQR 0.58 to 0.66), based on immediate and delayed memory (median AUPRC: 0.49, IQR: 0.43 to 0.52), and based on categorical verbal fluency (median AUPRC: 0.42, IQR 0.39 to 0.47). Every algorithm performed best when predicting poor cognition based on processing speed, working memory, and attention (median AUC ≥ 0.82 for each) compared to other subdomains. The lowest median AUCs and AUPRCs were observed when predicting poor cognition based on categorical fluency.

### Influence of wearable device features when predicting poor cognitive function

The contribution of each feature towards the final model predictions across the CatBoost, XGBoost, and Random Forest models was examined using SHapley Additive exPlanations (SHAP) plots. When predicting poor cognition based on processing speed, working memory, and attention we found that lower total activity variability (standard deviation), less time spent in moderate-vigorous activity per day, greater sleep efficiency variability, and lower maximum activity levels all contributed to higher risk predictions for poor cognition (Figure 4). The influence of activity levels on each model's risk predictions was similar when predicting poor cognition based on immediate and delayed memory (Figure 5) and categorical verbal fluency (Figure 6).

**Accelerometer-derived sleep and activity parameters are associated with multiple subdomains of cognition**

Results from association analyses showed that higher activity levels during non-sleep periods were univariably associated with better cognition across multiple measures of activity and all categorizations of cognitive function. The strength of associations was generally stronger when cognition was categorized based on processing speed, working memory, and attention or immediate and delayed memory compared to categorical verbal fluency (Figure 7). Higher levels of sedentary activity were associated with worse cognition across all subdomains in demographics-adjusted models, as was greater variability in L5 timing. In fully adjusted models, greater total activity variability was strongly associated ($p < 0.001$) with a lower odds of having poor cognition based on processing speed, working memory, and attention (per 1-sd increase, OR: 0.69, 95% CI: 0.60 to 0.79) as well as immediate and delayed memory (per 1-sd increase, OR: 0.78, 95% CI: 0.68 to 0.89). Greater sleep duration variability was also associated with a higher likelihood of having poor processing speed, working memory, and attention (per 1-sd increase, OR: 1.14, 95% CI: 1.02 to 1.27, p: 0.025). A complete description of all associations can be found in Figures 7 and 8 as well as Supplementary Tables 7 through 9.

**Associations between ambient light exposure and cognitive function vary across subdomains**

Additional association analyses examining the relationship between ambient light exposure captured using wearable device and cognition revealed varying relationships across cognitive subdomains. No single light exposure metric was associated with cognition across subdomains in the fully adjusted models (Figure 8). However, the kurtosis of an individual's light exposure distribution was significantly associated ($p < 0.05$) with an increased risk of poor cognition based on processing speed, working memory, and attention as well as immediate and delated memory. Higher light exposure variability (standard deviation) during non-sleep and M10 periods was associated with a decreased risk of poor cognition based on immediate and delayed memory as well as categorical verbal fluency.

## Discussion

In this study we found that the ability of machine learning models to assess cognition using wearable device data varied across cognitive subdomains. More specifically, we found that CatBoost, XGBoost, and Random Forest models exhibited better predictive capabilities when predicting poor cognitive function based on processing speed, working memory, and sustained attention compared to immediate and delayed memory as well as categorical verbal fluency. Our study provides proof-of-concept that wearable device data could be used to monitor cognitive capabilities among older adults, though the suitability of wearable-based monitoring approaches appears to vary across different measures of cognition. Significant relationships between accelerometer-derived parameters with multiple cognitive subdomains were also identified that could serve as targets for future causal studies seeking to understand how sleep, activity, and circadian rhythms influence cognitive function.

Each type of machine learning algorithm we evaluated performed best when identifying older adults with poor cognition based on processing speed, working memory, and attention as measured by the Digit Symbol Substitution Test (DSST). The DSST is well-suited to detect changes in cognitive function because it assesses multiple cognitive subdomains sensitive to aging and cognitive decline[19], thus, our study indicates that monitoring the processing speed, working memory, and sustained attention of older adults using wearable device data may be a viable alternative to existing monitoring approaches. However, our study additionally highlights the need for caution to be taken when leveraging wearable device data given that predictive ability varied considerably across different cognitive subdomains. Existing evidence supports this call for caution as similar variability was observed in a wearable-based cognition prediction study of older adults in Singapore[17] with Mild Cognitive Impairments (MCI). As such, careful consideration will be needed when designing wearable-based cognitive monitoring systems to ensure the target variables in question can be predicted accurately using wearable device data.

A central insight uncovered in this study is the apparent relationship between the standard deviation of total activity levels (hereafter activity variability) and cognitive function among older adults. In association analyses total activity variability exhibited a stronger relationship with cognition than average total activity across all categorizations of poor cognition (Figure 7). In the predictive analyses SHAP plots showed that higher activity variability resulted in lower risk predictions for poor cognition (Figures 4-6) in the XGBoost, CatBoost, and Random Forest models across all cognitive subdomains. Interestingly, activity variability was highly correlated with average total activity (Pearson's r: 0.90), yet activity variability was favored during feature selection for the prediction models (Figure 2). Instead of average total activity, average time spent in moderate-to-vigorous activity was favored during feature selection, indicating that intensity-specific activity averages are more useful for monitoring cognitive function than overall averages. Indeed, intensity-specific activity averages generally showed a stronger association with cognition compared to overall activity averages in our study as well. Future longitudinal accelerometer studies may wish to examine if changes in activity variability predate cognitive decline to determine if activity variability could be used as a marker for future impairments.

Several other activity variability metrics were associated with poor cognition in this study, namely day-to-day variability in moderate-to-vigorous activity, sedentary activity, and M10 activity. Greater values of each metric were univariably associated with a lower risk of poor cognitive function across all cognitive subdomains. However, more time spent in sedentary activity was associated with a greater risk of poor cognition, representing a case where the relationships between mean and variability metrics with cognition were in opposing directions. A wealth of evidence has already linked sedentary activity to greater risks of adverse cognitive outcomes[11,20,21], while this study is the first to show older adults with greater sedentary activity variability are less likely to have poor cognition. Our findings likely stem from older adults who are more active on average having greater levels of activity variability. Older adults who are more physically capable have the freedom to pursue activities of varying intensities, whereas less active and less physically capable older adults are bound to more sedentary behaviors. Previous studies have shown that average activity levels are lower among older adults years prior to the development of cognitive impairments and continue to decline as impairments intensify[11,20,21], but future studies are needed to determine if changes in activity variability across different levels of intensity follow in-step with declines in average activity.

Measures of ambient light exposure showed limited and varied relationships with cognition in this study, adding to the mixed evidence present in the literature. Some evidence points towards greater outdoor sun exposure having positive associations with cognition[22]. Other findings have shown that light from technological devices such as TVs negatively impact sleep quality[23] and by extension may have negative downstream effects on cognition. As such, the type and timing of light exposure seem to be more important than overall exposure. The ambient light metrics examined in this study could not differentiate between sun exposure and non-sunlight exposure, though we did examine exposure during sleep, non-sleep, M10, and L5 periods. In fully adjusted models, greater variability in ambient light exposure during non-sleep and M10 periods were associated with lower odds of having poor cognition based on immediate and delayed memory as well as categorical verbal fluency. Again, we observed a high correlation between average and variability metrics for M10 light exposure (Pearson's r: 0.78) and non-sleep light exposure (Pearson's r: 0.77) yet neither average metric was associated with any categorization of poor cognition in fully adjusted models. Reasons for this discrepancy are unclear, and additional studies are needed to determine if the findings in this study generalize.

Our study has several strengths. It is, to our knowledge, the first examine the feasibility of using machine learning models and wearable device data to differentiate between older adults with normal and poor cognition based on categorizations derived from multiple cognitive tests. Our work holds implications for the potential viability of developing wearable-based cognitive monitoring systems, primarily by demonstrating that wearable device data and machine learning can accurately identify older adults with poor cognition as measured by processing speed, working memory, and sustained attention. We further derived novel accelerometer and ambient light exposure metrics for use in both our prediction models and associative analyses. Previously unidentified relationships were established between metrics derived from wearable device data with cognitive function, namely those relating to variability in sleep parameters, activity levels, and ambient light exposure. Such metrics could serve as targets for future causal studies seeking to identify early markers of cognitive decline or additional association analyses seeking to gain a better understanding of the interplay between sleep, activity, and light exposure metrics with cognition among older adults.

This study is not without limitations. First, the cross-sectional nature of the data meant we could not attempt to detect changes in cognition across time and were limited to predicting current cognitive function at a single timepoint. Cross-sectional data also made examining causal relationships infeasible. Our machine learning models were also limited by the data that was collected by the NHANES wearable devices. There is emerging evidence that physiological signals such as skin temperature and heart rate variability are associated with cognitive function[24-26] and are useful for cognition prediction tasks[17,27]. While some commercially available wearables have the capability to collect such information, the devices used in the NHANES do not. As such, quantifying the full potential of wearable device data to monitor cognition may best be undertaken using wearable devices that can collect data beyond acceleration and ambient light exposure.

## Methods

**Data Sources and Study Design**
The National Health and Nutrition Examination Survey[18] (NHANES) is a nationally representative survey of non-institutionalized Americans aged six and older that has been conducted on a yearly basis since 1999. The NHANES utilizes a multistage probability sampling design[18] and collects data from roughly 5000 Americans each year. In the 2011-14 NHANES waves participants were asked to wear a wrist-worn physical activity monitor with an embedded accelerometer as much as possible over nine consecutive days, after which the activity data was made available in Monitor-Independent Movement Summary (MIMS[28]) units. The device also captured ambient light exposure which was made available in minute-level summaries. Over 16,000 NHANES participants wore accelerometers between 2011-2014, and participants over the age of 60 were additionally administered three cognitive tests: the Consortium to Establish a Registry for Alzheimer's Disease Word-Learning subtest (CERAD-WL), Animal Fluency Test (AFT), and Digit Symbol Substitution Test (DSST)[29]. NHANES received ethics approval from the National Center for Health Statistics Research Ethics Review Board and all participants gave their informed consent. The handling, processing, and analysis of all NHANES data in this study was done in accordance with data access and ethical regulations stipulated by the National Center for Health Statistics.

In this study we used the 2011-2014 NHANES waves. Our primary objective was to determine if machine learning models trained on wearable device data can differentiate between poor cognition and normal cognition across different subdomains of cognitive function. We further aimed to quantify associations between sleep, activity, circadian, and ambient light exposure metrics with cognition. Participants were excluded if they were under the age of 60, had missingness in their cognitive test data, had less than five days of accelerometer wear time, or if they had less than three nights of valid sleep data. This study is reported following the transparent reporting of a multivariable prediction model for individual prognosis or diagnosis (TRIPOD) statement[30] where appropriate.

**Cognitive Measures**
The NHANES includes three cognitive tests: the Consortium to Establish a Registry for Alzheimer's Disease Word-Learning subtest (CERAD-WL), which tests immediate and delayed memory, the Animal Fluency Test (AFT), which assesses executive function by quantifying categorical verbal fluency, and the Digit Symbol Substitution Test (DSST), which tests working memory, processing speed, and sustained attention[29]. The CERAD-WL consists of three rounds of immediate word recall and one round of delayed recall. Ten unrelated words are tested in each round with aggregate scores from all four rounds ranging from zero to forty. During the AFT participants are given one point per animal named in a one-minute period with no upper bound on the maximum number of points. For the DSST each participant is first given a key matching nine numbers to nine symbols. Afterwards, each person is assessed based on how many symbols can be written below 133 numbers based on the provided key in two minutes with total scores ranging from 0 to 133. For all three tests higher scores indicate better cognitive function, and each test has been used in previous studies examining cognition among older adults[31-34]. Previously, the Center for Disease Control's National Center for Health Statistics has used the 25th quantile as a cutoff for low cognitive performance on each test[29]. As such, poor cognition was categorized using the 25th quantile as a cutoff (20 for the CERAD-WL, 33 for the DSST, and 12 for the AFT) in this study for each test.

**Feature Derivation**

Activity and sleep features were derived from the NHANES accelerometer data. We determined sleep onset and offset times using a publicly available unsupervised Hidden Markov Model (HMM) developed for sleep-wake identification[35]. The HMM algorithm has previously been validated against gold standard polysomnography and has a sensitivity of over 99% when inferring if a person is sleeping at a given point in time. Four sleep-related metrics were calculated using the HMM algorithm: sleep onset times, sleep offset times, sleep duration, and sleep efficiency. Sleep duration measures the total time spent sleeping whereas sleep efficiency is the ratio of sleep duration over the total time between initial sleep onset and final sleep offset. The average and standard deviation of each sleep metric across all nights of valid sleep data were included as features in this study. To account for the cyclic nature of the time data we used the circular mean and standard deviation[36] instead of the arithmetic equivalents when deriving sleep onset and offset features. Next, we calculated each person's total activity levels across sleep and non-sleep periods, after which the average and standard deviations were calculated and included as features. Activity categorizations were then made where each minute of non-sleep accelerometer data was classified as sedentary (MIMS <15.9), light (MIMS between 15.9 and 19.6, inclusive), or moderate-to-vigorous (MIMS >19.6) based on validated cutoffs from prior work[37]. The average and standard deviation of the proportion of daily time spent in each activity category were then derived and included as features. Non-wear observations were excluded prior to the derivation of the activity metrics.

The second type of wearable device features were derived using statistical and signal processing techniques[38]. We first calculated several summary statistic features from each participant's accelerometer data excluding non-wear observations: the mean, median, standard deviation, maximum, minimum, 10th quantile, 25th quantile, 75th quantile, 90th quantile, skewness, kurtosis, and differential entropy. A Fast Fourier Transform (FFT) was then applied to each participant's acceleration data, from which we used the fifteen frequencies corresponding to the largest fifteen amplitude values as features.

Features related to circadian rhythm analyses were also included in this study[39]. We first identified the hour in the middle of each person's least active five-hour (L5) period and most active ten-hour period (M10) each day. The circular average and standard deviation of the L5 and M10 midpoints across the wear period were included as features. Next, we calculated the average and standard deviation of individual activity levels during the L5 and M10 periods. The mean and standard deviation of each person's relative amplitude (RA) and intraday variability (IV) for their activity data were also derived along with inter-daily stability (IS) of activity levels across the wear period. Finally, we gathered each participant's 24-, 12-, and 8-hour FFT signal strengths as the final features.

Lastly, features were derived based on each participant's ambient light exposure, which was also collected by the wearable device in the NHANES waves used in this study. The mean and standard deviation of ambient light exposure during sleep, non-sleep, L5, and M10 period were derived, as were all statistical and signal processing features described previously using each participant's complete ambient light exposure data.

**Feature Selection and Prediction Models**

Multiple models were developed to predict poor cognition based on three categorizations spanning multiple cognitive subdomains. One categorization was based on processing speed, working memory, and attention (DSST scores), another based on immediate and delayed memory (CERAD-WL scores), and lastly one based on categorical verbal fluency (AFT scores) respectively. As such, we conducted separate feature selection procedures for all three prediction tasks. The features were selected from a candidate set that included age, education, and all previously described accelerometer derived and ambient light exposure features. Age and education were included because of their known associations with cognitive function, relative ease to collect through an interface on a smart device, and because both features would not require manual updating for repeated cognitive assessments. For comparison purposes we additionally developed models that only included age and education (Supplementary Table 4).

Features for the models were chosen using a forward-selection procedure and CatBoost[40], a gradient boosting machine learning algorithm. Gradient boosting algorithms have been shown effective in previous clinical prediction studies[41,42], can capture complex nonlinear interactions between features, and provide robust results in the presence of multicollinearity. The features were first ranked from most to least important based on their mean absolute SHapley Additive exPlanation (SHAP) values[43] using a CatBoost model containing all candidate features. SHAP values enable model interpretability by quantifying the marginal contribution of a given feature to a model's final predictions, thereby unveiling both the feature's importance and how specific values of that feature influence the model's output. From there, we recursively trained CatBoost models adding one feature at a time from most to least important. Each model was evaluated through 10-fold cross-validation based on the Area Under the Curve (AUC). The final feature sets were chosen subjectively based on when no reasonable improvements in AUC were observed across the feature selection procedure.

After the features were selected we developed nine models: one CatBoost, one XGBoost[44], and one Random Forest model for predicting poor cognition based all three categorizations of poor cognitive function. Hyperparameters for each model were tuned through 10-fold cross validation using Optuna, an automated hyperparameter optimization framework[45]. The specifics of the tuning procedure for a single model are as follows. First, we defined a range of possible values for each hyperparameter and chose the Area Under the receiver operating characteristic Curve (AUC) as the performance metric for model evaluation. Next, we ran 200 Optuna trials to find the optimal hyperparameter values. For each trial Optuna suggests hyperparameter values within the prespecified range, a model with the suggested hyperparameters is put through 10-fold cross validation, and the average AUC is calculated from the validation fold predictions. Over many trials the optimization framework learns to suggest hyperparameter values that improve the performance of the model, in our case by maximizing the AUC. For the CatBoost and XGBoost models we tuned the learning rate, number of trees, tree depth, and subsampling done by each tree. For the Random Forest models we tuned the tree depth and maximum number of trees.

**Predictive Modelling**

After the final features and hyperparameters were chosen we evaluated each model through 20-repeats of 10-fold cross validation, which has been recommended to obtain optimism adjusted performance metrics for clinical prediction models[46]. Models were assessed based on the Area Under the receiver operating characteristic curve (AUC) and the Area Under the Precision Recall Curve (AUPRC). The AUC was chosen as it is a commonly used metric in clinical prediction studies that quantifies the ability of a model to differentiate between two classes, in this case normal versus poor cognition. The AUPRC was further included given the imbalanced nature of the data as it captures the ability of a model to predict a positive minority class, in our case poor cognitive function. AUC scores greater than 0.5 indicate better-than-random predictions whereas AUPRC scores larger than the proportion of the minority class indicate better-than-random predictions. Boxplots were constructed for each model from the 200-validation set performance metrics obtained during repeated cross-validation which were also reported using the median and interquartile range (IQR). During repeat cross-validation missing values were imputed using K-Nearest Neighbors imputation[47] with K=1 on each training and validation set separately. SHAP plots were generated for each of the final models to quantify feature importance and enable explainable predictions. Lastly, a sensitivity analysis was performed where the repeated cross-validation procedure was replicated using complete case data (Supplementary Table 5).

**Measuring Associations**

Associations were examined between accelerometer-derived and ambient light exposure features with each cognitive subdomain. Relevant confounders were gathered for inclusion in adjusted models, namely age, sex, education, household income, marital status, diabetic status, arthritis, depression, whether someone smokes, whether someone drinks alcohol, and heart disease. Depression was derived from responses to the nine-item Patient Health Questionnaire[48] (PHQ-9), for which participants with scores greater than 10 were categorized as depressed. Participants were categorized as either current or non-current smokers and drinkers, while arthritis, diabetes, and heart disease were determined using self-reported data.

Associations were examined using complete case data between all accelerometer and ambient light features in univariable, demographic, and fully adjusted models. The demographic models included age, sex, education, household income, and marital status while the fully adjusted models further included depression, diabetes, arthritis, heart disease, whether someone smokes, and whether someone drinks. Odds ratios were reported per one standard deviation increase.

# Data Availability

The data analyzed during this study are all publicly available through the National Health and Nutrition Examination Survey's website at https://wwwn.cdc.gov/nchs/nhanes/

### Code Availability
The underlying code for this study is available on GitHub and can be accessed via this link [*to be provided upon acceptance or at the editor's request prior to*].

### Acknowledgements
This study was funded by City University of Hong Kong, Hong Kong SAR, China internal research grant #9610473. The funder played no role in study design, data collection, analysis and interpretation of data, or the writing of this manuscript.

### Author Contributions
CS designed the study with supervision from XL and JL. CS and TL performed the statistical analyses with oversight from XL. CS, XL, and JL contributed to writing and editing the manuscript. All authors read and approved the final manuscript.

### Competing Interests
All authors declare no financial or non-financial competing interests.

# Figures

Table 1. Cohort Characteristics

| Characteristic | Mean (SD) or N (%)[a] |
|---|---|
| Age | 69.55 (6.77) |
| Sex (Male) | 1282 (51.71) |
| Education | |
|     Less than High School | 609 (24.57) |
|     High School | 597 (24.08) |
|     Some college or Associates Degree | 716 (28.88) |
|     College or above | 555 (22.39) |
| Marital status | |
|     Married or Living w/ Partner | 1421 (57.32) |
|     Divorced or Separated | 420 (16.94) |
|     Widowed | 492 (19.85) |
|     Never married | 144 (5.81) |
| Household income (1000s of US Dollars) | |
|     <45k | 1341 (54.09) |
|     45k to 100k | 609 (24.57) |
|     >100k | 321 (12.95) |
| Diabetic | 592 (23.88) |
| Depressed | 190 (7.66) |
| Smoker | 304 (12.26) |
| Drinks Alcohol | 1218 (49.13) |
| Heart Disease | 240 (9.68) |
| Arthritis | 1240 (50.02) |
| Poor processing speed, working memory, and attention | 625 (25.21) |
| Poor immediate and delayed memory | 609 (24.57) |
| Poor categorical verbal fluency | 587 (23.68) |

[a]Categorical counts not summing to the total sample size N= 2479 indicate missingness. There were no missing values for age and sex. A full description of missingness across the data can be found in Supplementary Table 6

Figure 1. Hourly activity averages stratified by cognitive status

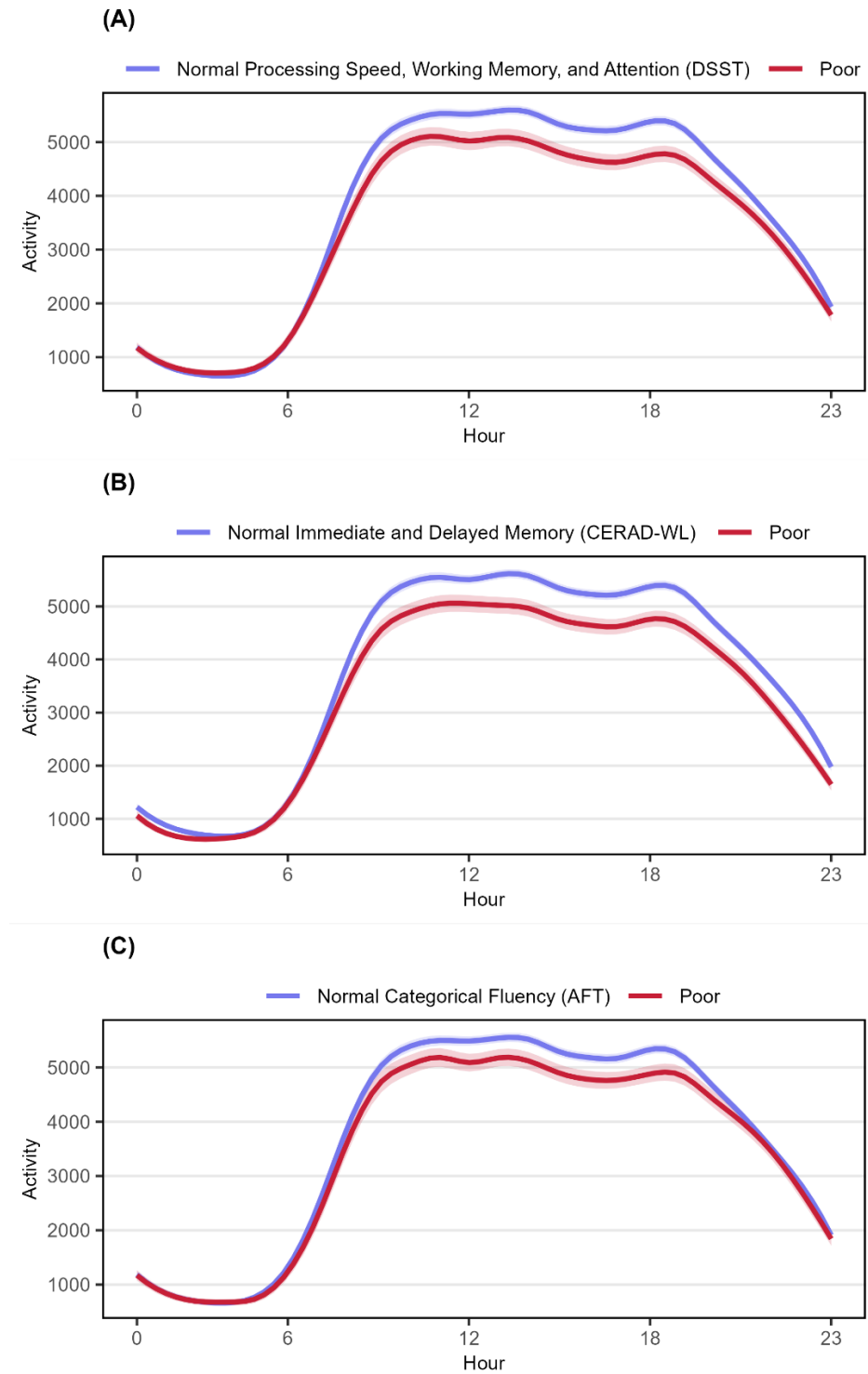

The plot shows hourly activity levels stratified by poor cognitive function based on (A) processing speed, working memory, and attention as measured by the DSST (B) immediate and delayed memory as measured by the CERAD-WL and (C) categorical verbal fluency as measured by the AFT. Error bars represent 95% confidence intervals calculated for each hour of activity where hourly averages were connected using spline interpolation.

Figure 2. Feature selection results

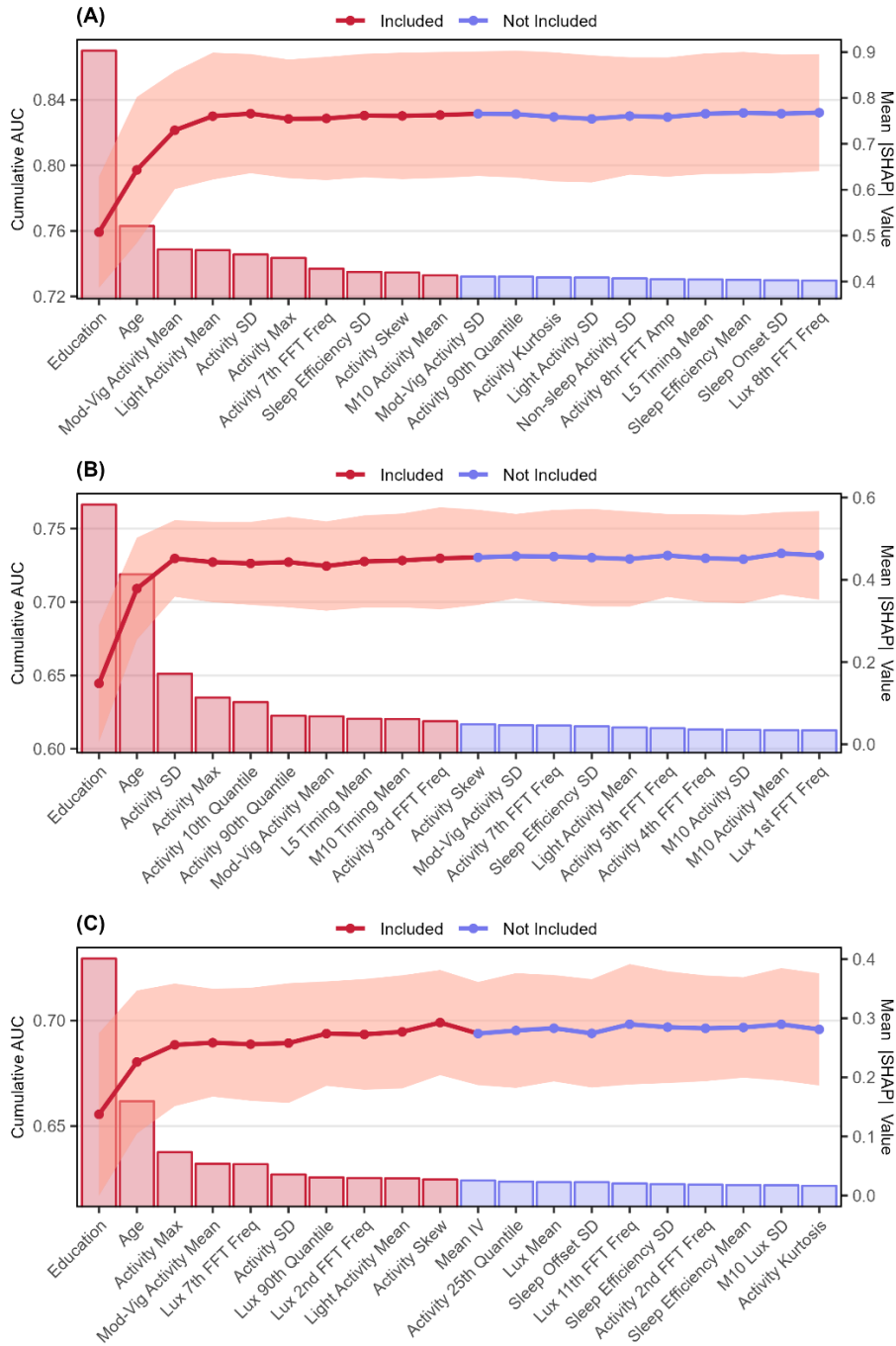

The figure shows the results from the forward feature selection procedure for predicting poor cognitive function based on (A) processing speed, working memory, and attention as measured by the DSST (B) immediate and delayed memory as measured by the CERAD-WL and (C) categorical verbal fluency as measured by the AFT. Bars represent the mean absolute SHAP value for a given feature (right y-axis) and points represent mean validation set AUC from 10-fold cross-validation (left y-axis). Error bars are the mean AUC ± 1 standard deviation. Only the first 20 features are shown. Figures showing the selection procedure across all candidate features can be found in Supplement Figure 1.

Figure 3. Repeated cross-validation results

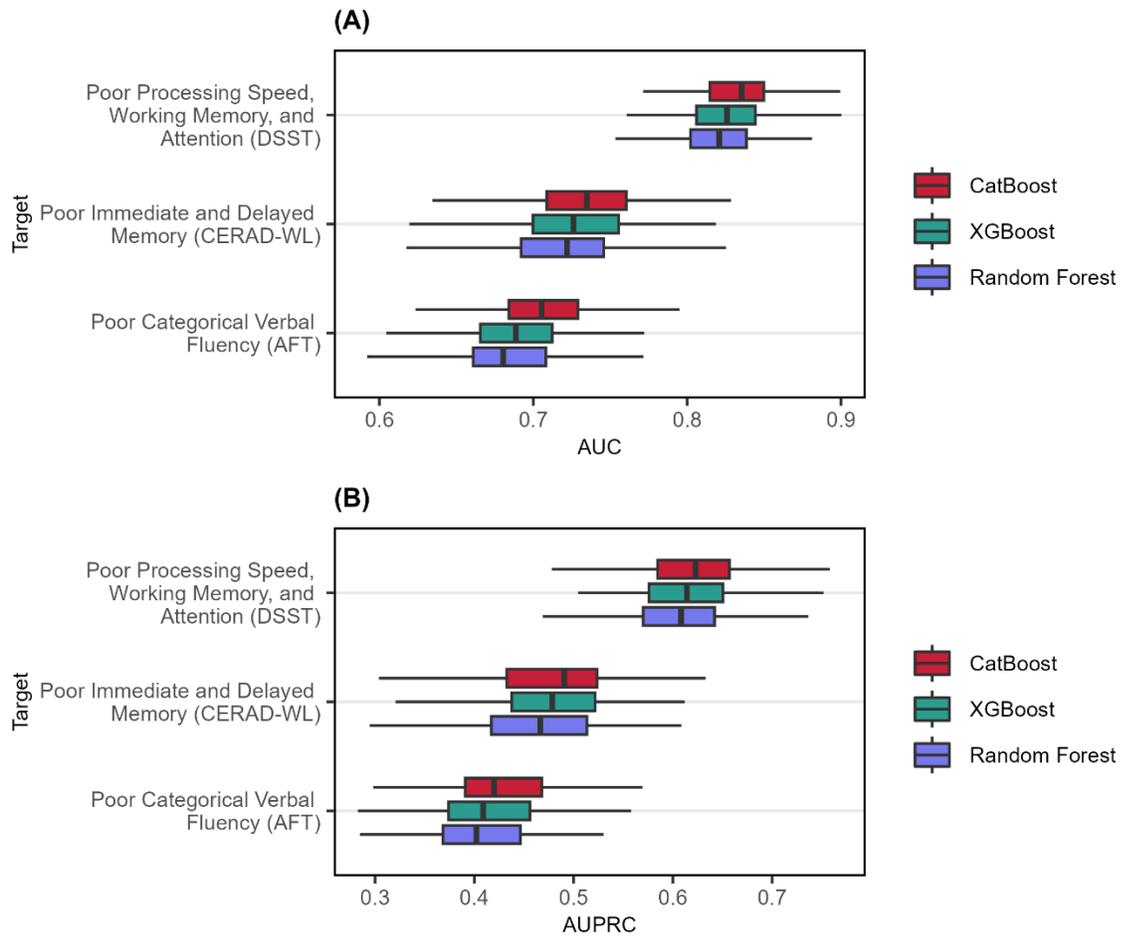

The plot indicates the predictive performance of each model type as evaluated by (A) the AUC and (B) the AUPRC through 20 repeats of 10-fold cross-validation.

Table 2. Repeated cross-validation results

|  | Median [IQR] | | |
| --- | --- | --- | --- |
| Target / Model Type | Poor processing speed, working memory, and attention (DSST) | Poor immediate and delayed memory (CERAD-WL) | Poor categorical verbal fluency (AFT) |
| CatBoost | | | |
|     AUC | 0.84 [0.81, 0.85] | 0.73 [0.71, 0.76] | 0.71 [0.68, 0.73] |
|     AUPRC | 0.62 [0.58, 0.66] | 0.49 [0.43, 0.52] | 0.42 [0.39, 0.47] |
| XGBoost | | | |
|     AUC | 0.83 [0.81, 0.84] | 0.73 [0.7, 0.76] | 0.69 [0.67, 0.71] |
|     AUPRC | 0.61 [0.58, 0.65] | 0.48 [0.44, 0.52] | 0.41 [0.37, 0.46] |
| Random Forest | | | |
|     AUC | 0.82 [0.8, 0.84] | 0.72 [0.69, 0.75] | 0.68 [0.66, 0.71] |
|     AUPRC | 0.61 [0.57, 0.64] | 0.47 [0.42, 0.51] | 0.4 [0.37, 0.45] |

Figure 4. SHAP plots for predicting poor processing speed, working memory, and attention

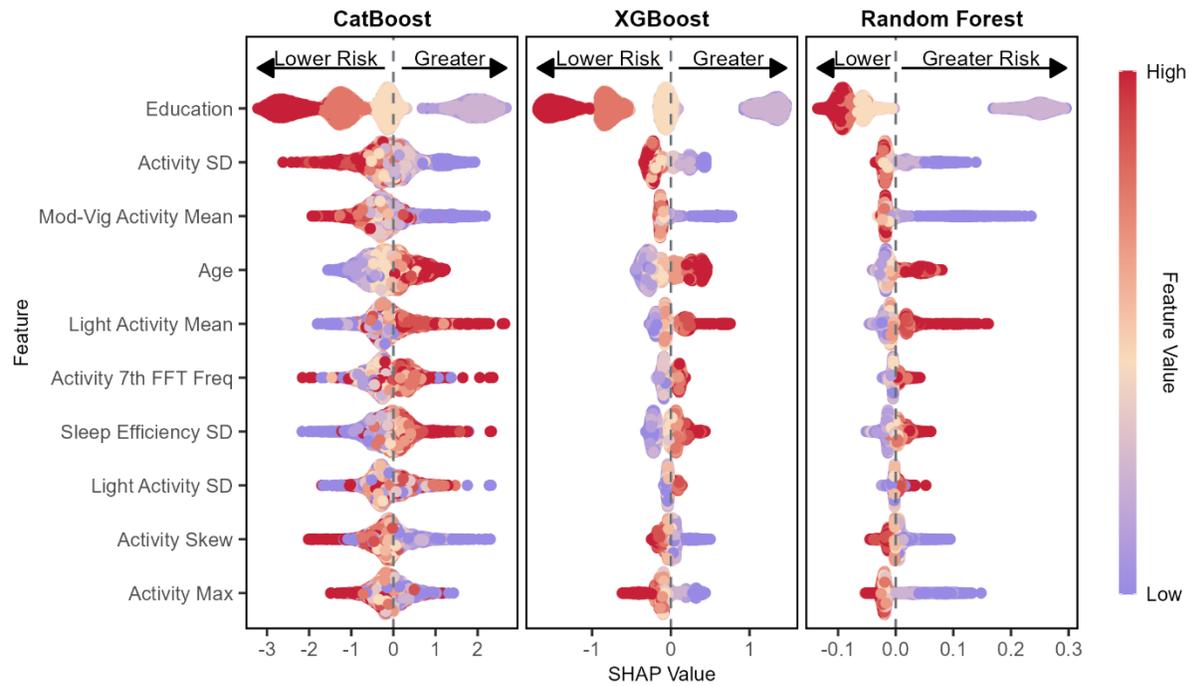

The figure shows the SHAP values for predicting poor processing speed, working memory, and attention as measured by the DSST for the CatBoost model, the XGBoost model, and the Random Forest model. Features are ordered top to bottom from most to least important based on the mean absolute SHAP value of the CatBoost model.

Figure 5. SHAP plots for predicting poor immediate and delayed memory

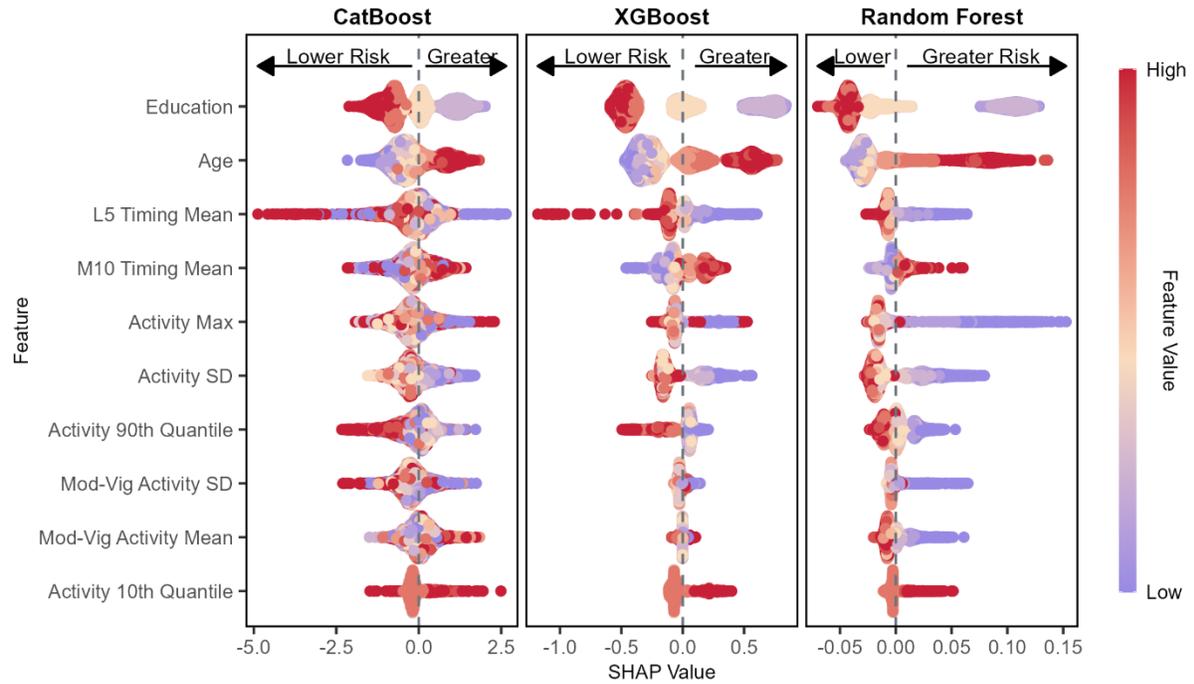

The figure shows the SHAP values for predicting poor immediate and delayed memory as measured by the CERAD-WL for the CatBoost model, the XGBoost model, and the Random Forest model. Features are ordered top to bottom from most to least important based on the mean absolute SHAP value of the CatBoost model.

Figure 6. SHAP plots for predicting poor categorical verbal fluency

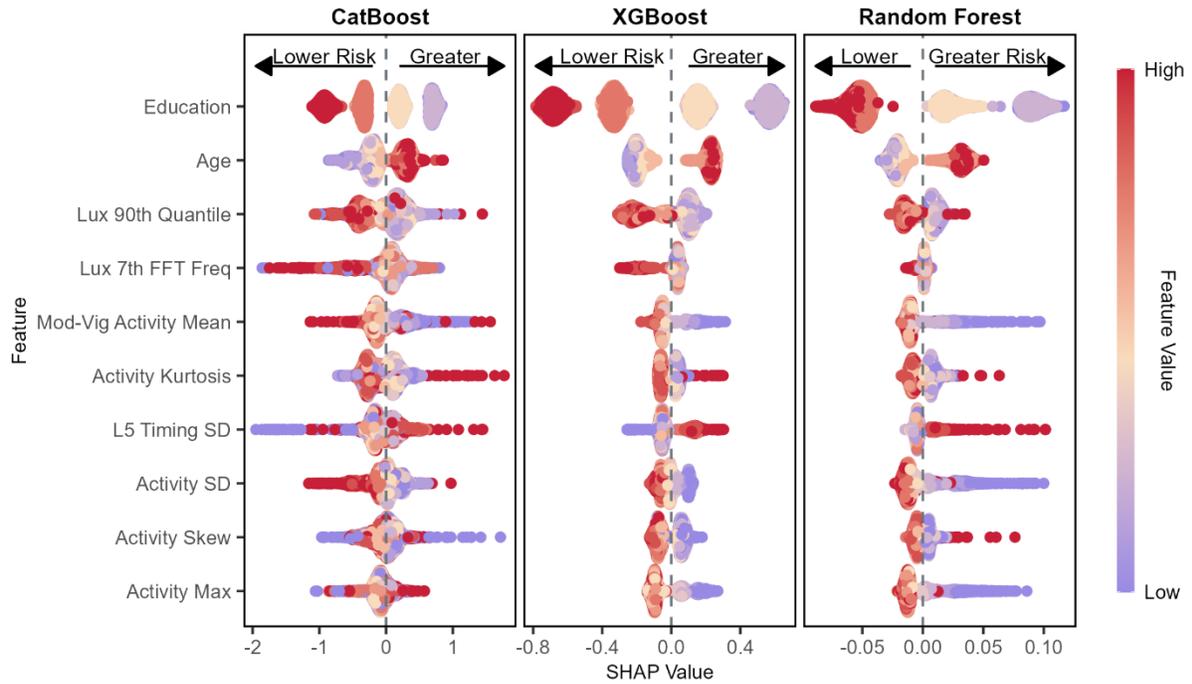

The figure shows the SHAP values for predicting poor categorical verbal fluency as measured by the AFT for the CatBoost model, the XGBoost model, and the Random Forest model. Features are ordered top to bottom from most to least important based on the mean absolute SHAP value of the CatBoost model.

Figure 7. Associations of activity and sleep metrics with cognitive function

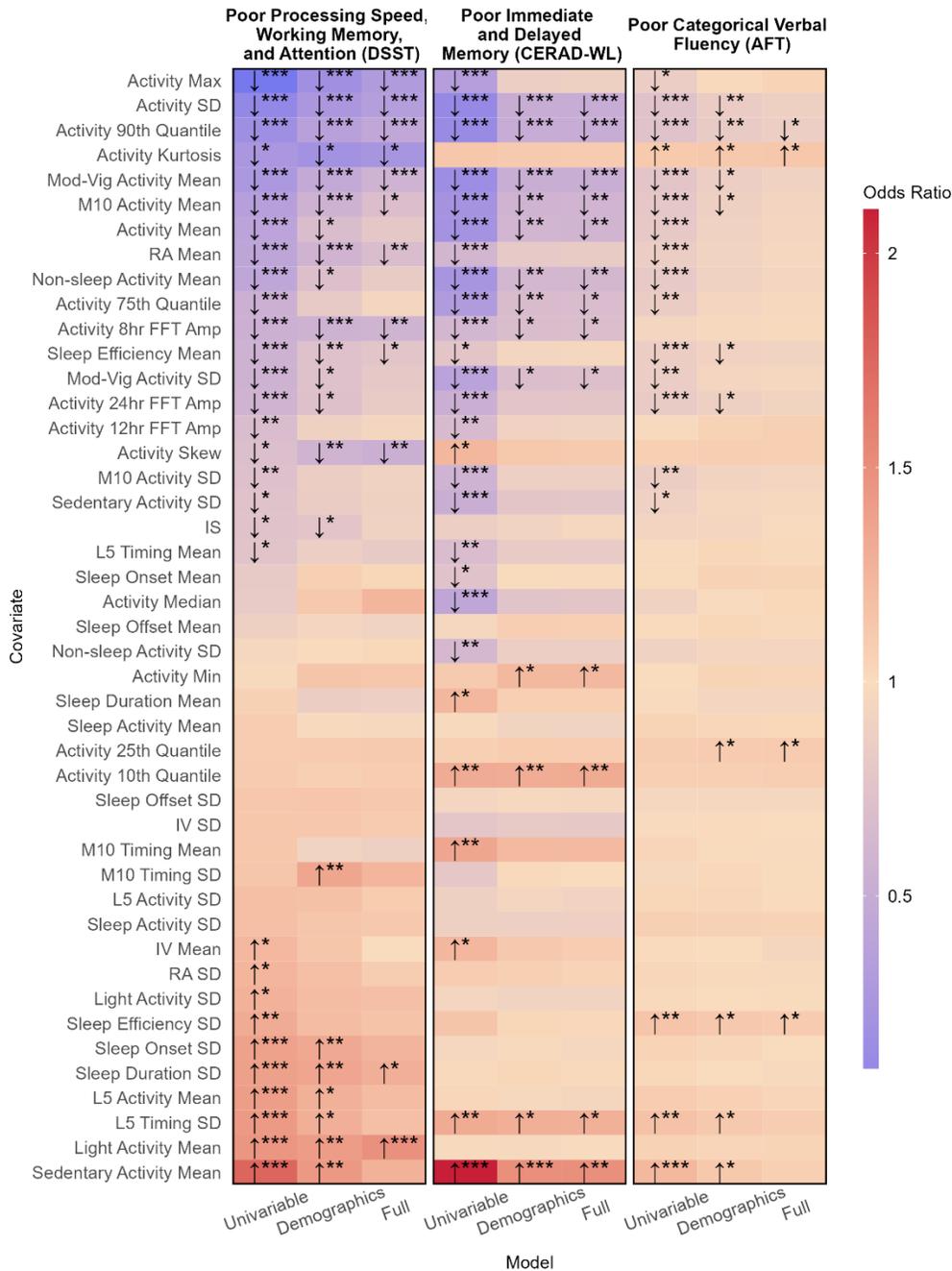

This figures indicates the strength of association between accelerometer derived activity, sleep, and circadian metrics with poor processing speed, working memory, and attention (leftmost panel), poor immediate and delayed memory (middle panel), and poor categorical verbal fluency (rightmost panel). Arrows indicate whether larger values for a metric are associated with a greater (odds ratio > 1, upward facing arrow) risk of having poor cognitive function or lower (odds ratio < 1, downward facing arrow) risk of having poor cognitive function. Stars represent the degree of statistical significance (* < 0.05, ** < 0.01, *** < 0.001). Tiles without arrows and stars had no significant associations. Odds ratios are reported per one standard deviation increase.

Figure 8. Associations of ambient light exposure metrics with cognitive function

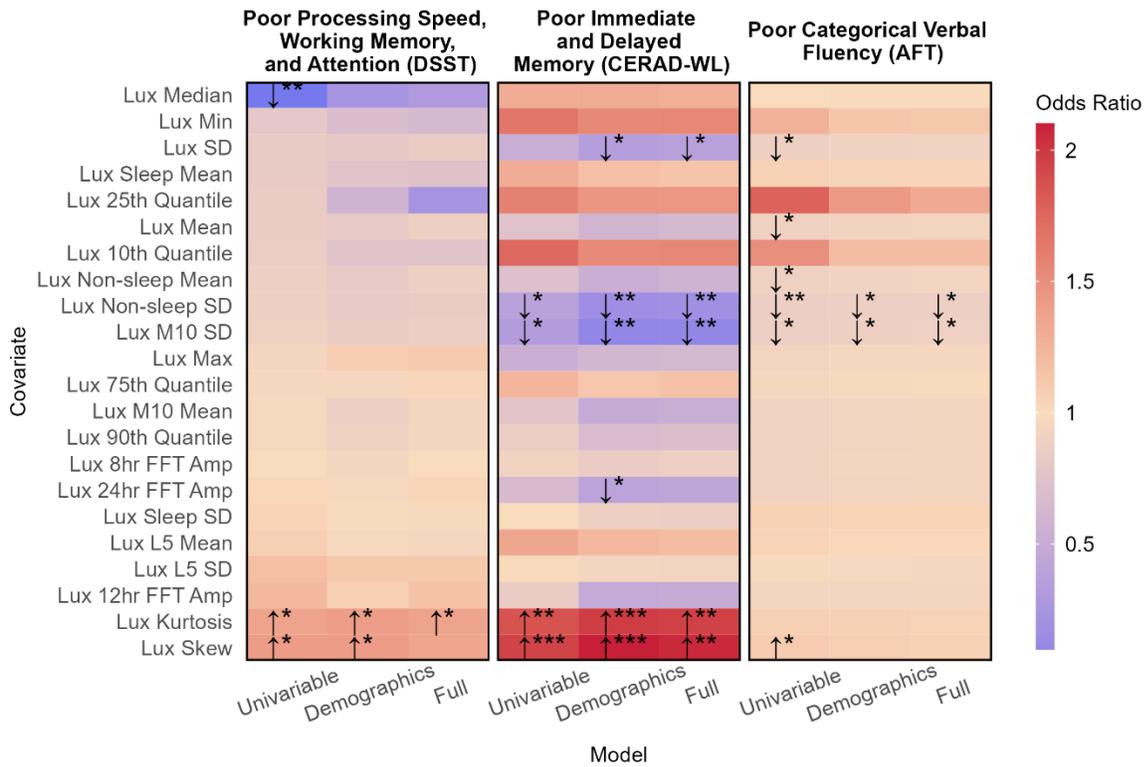

This figure indicates the strength of association between ambient light exposure with poor processing speed, working memory, and attention (leftmost panel), poor immediate and delayed memory (middle panel), and poor categorical verbal fluency (rightmost panel). Arrows indicate whether larger values for an accelerometer metric are associated with a greater (odds ratio > 1, upward facing arrow) risk of having poor cognitive function or lower (odds ratio < 1, downward facing arrow) risk of having poor cognitive function. Stars represent the degree of statistical significance (* < 0.05, ** < 0.01, *** < 0.001). Tiles without arrows and stars had no significant association. Odds ratios are reported per one standard deviation increase.